# The Exciting Star of the Berkeley 59/Cepheus OB4 Complex and Other Chance Variable Star Discoveries

Daniel J. Majaess[1] and David G. Turner[1]
David J. Lane[2]
Kathleen E. Moncrieff
*Department of Astronomy and Physics, Saint Mary's University, Halifax, NS, B3H 3C3, Canada*

[1] Visiting Astronomer, Dominion Astrophysical Observatory, Herzberg Institute of Astrophysics, National Research Council of Canada.
[2] Abbey Ridge Observatory, Stillwater Lake, NS, Canada.

**Abstract** A study is presented regarding the nature of several variable stars sampled during a campaign of photometric monitoring from the Abbey Ridge Observatory: 3 eclipsing binaries, 2 semiregulars, a luminous Be star, and a star of uncertain classification. For one of the eclipsing systems, BD+66°1673, spectroscopic observations reveal it to be an O5 V((f))n star and the probable ionizing star of the Berkeley 59/Cep OB4 complex. An analysis of spectroscopic observations and *BV* photometry for Berkeley 59 members in conjunction with published observations imply a cluster age of ~2 Myr, a distance of $d = 883 \pm 43$ pc, and a reddening of $E_{B-V} = 1.38 \pm 0.02$. Two of the eclipsing systems are Algol-type, but one appears to be a cataclysmic variable associated with an X-ray source. ALS 10588, a B3 IVn star associated with the Cepheid SV Vul, is of uncertain classification, although consideration is given to it being a slowly pulsating B star. The environmental context of the variables is examined using spectroscopic parallax, 2MASS photometry, and proper motion data, the latter to evaluate the membership of the variable B2 Iabe star HDE 229059 in Berkeley 87, an open cluster that could offer a unique opportunity to constrain empirically the evolutionary lineage of young massive stars. Also presented are our null results for observations of a sample of northern stars listed as Cepheid candidates in the Catalogue of Newly Suspected Variables.

## 1. Introduction

The present study is the result of a survey of variable stars initiated at the Abbey Ridge Observatory (ARO). Most of the discoveries resulted from the variability of potential reference and check stars in the fields of Cepheid variables, the campaign's primary objective being to establish period changes for northern hemisphere Cepheids (Turner 1998, Turner *et al*. 1999). In two cases the interest lies in cluster stars discovered to be variable.

The program began in 1996 at the Burke-Gaffney Observatory (BGO) of Saint Mary's University, and was recently transferred to the ARO, which is located at a darker site. The stability of the ARO site and an upgrading of the photometric reduction routines facilitated the discoveries, some of which are described here. A preliminary survey was also made of stars in the *Catalogue of Newly Suspected Variables* (NSV, Samus *et al*. 2004) thought to be potential Cepheid variables, with the goal of expanding the Galactic sample and eventually studying their period changes. Rate of period change, in conjunction with light amplitude, has been demonstrated to be an invaluable parameter for constraining a Cepheid's crossing mode and likely location within the instability strip (Turner *et al*. 2006). Such constraints permit further deductions to be made concerning a Cepheid's intrinsic reddening and pulsation mode, and offer yet another diagnostic tool for establishing Cepheids as members of open clusters, such calibrators being the foundation for the extragalactic distance scale (Turner and Burke 2002).

The photometric signatures of some variable stars can be sufficiently ambiguous that spectroscopic follow-up was necessary to resolve the true nature of the light variations. Preliminary results for the best-studied of the variables, summarized in Table 1, are presented here in order of increasing right ascension. The results for the NSV stars are given at the end.

## 2. Observations and Equipment

The ARO is located in Stillwater Lake, a community ~23 km west of Halifax, Nova Scotia, Canada. The ARO allocates ~2 hours of observing time on each clear night to variable star research, with much of the remainder being used to search for extragalactic supernovae as part of the Puckett Observatory Supernova Search program. The site is quite dark, with typical Sky Quality Meter readings of 20.6 $V$ mag. arcsec$^{-2}$ on good nights. The observatory houses a 28-cm Celestron Schmidt-Cassegrain telescope equipped with an SBIG ST9-CCD camera and Bessel $B$ and $V$ filters. The facility is remotely accessible and completely automated, allowing unattended acquisition of astronomical and calibration images and providing a software pipeline that calibrates, combines, and performs differential aperture photometry. Much of the design and software development to realize the ARO's capability was completed in the form of the *Abbey Ridge Auto-Pilot* software (Lane 2007). Other software, in particular *MaxIm DL/CCD* (George 2007), provides many of the low-level image acquisition and processing functions.

Observations and image processing are guided by two data files provided by the observer. The first contains relevant information about each field to be imaged, including a unique identifier, center equatorial co-ordinates, and exposure times and number of exposures to be taken in each filter. If aperture photometry is to be performed automatically on the field, additional information is needed, including aperture photometry parameters, magnitude of the designated reference star, and equatorial co-ordinates for each star to be measured. The second file type contains a list of target fields (identifiers) to be observed on a given night.

All resulting photometric data represent means for multiple (15–25, or more) short-exposure images, typically of 1 to 60 seconds duration, taken in immediate succession, combined using a noise-reduction algorithm developed by DJL. The algorithm first calibrates the individual images instrumentally, then registers them spatially using stars in the field. The mean and standard deviation are computed for each pixel position in the stack of images. Pixels on a given image that deviate more than a specified number of standard deviations from the mean of pixels at the same pixel positions on the other images are rejected, and a new mean is computed. The process is iterated up to five times until all deviant pixels are rejected, although no more than 30% of the pixels at a given pixel position are ever rejected. The resulting mean, without inclusion of rejected pixels, is used to form the combined image, which is plate-solved astrometrically using the *PinPoint Astrometric Engine* software (Denny 2007).

Differential aperture photometry is performed on each combined image using initial aperture and sky annulus parameters, and equatorial co-ordinates of the primary target star, reference star, and any number of "check" or other stars. The sky annulus radius and size are pre-selected for each field to be appropriate for all stars measured. The equatorial position of each star is converted to its corresponding X and Y pixel position using the plate solution embedded in the fts header. Aperture photometry is performed on the reference star 3 times iteratively to determine its precise X and Y pixel position and full-width half maximum (FWHM) seeing disk. A new aperture size is set to a pre-determined multiple of the measured FWHM, and the sky annulus position is adjusted so that it remains at the original radius from the star's center. The new aperture parameters and the same technique are used for the remaining stars measured. The instrumental magnitude derived for the reference star is used to convert instrumental magnitudes for the remaining stars into differential magnitudes.

The output for each measured star includes pertinent information, such as FWHM, sky annulus background, signal-to-noise ratio, and maximum pixel value, all of which are invaluable when assessing the data quality. In most instances the reference stars are selected from the comprehensive Tycho-2 catalog (Høg *et al.* 2000), so there may be zero-point offsets, given that the reference stars are non-standards and the photometry is not explicitly normalized to the Johnson system. Also, differences in color between the reference and target stars are not normally taken into account.

The spectra used to classify the variables were obtained during two observing runs with the Dominion Astrophysical Observatory's 1.8m Plaskett telescope. During the first run in October 2006, the SITe-2 CCD detector was used to image spectra at a resolution of 120 Å mm$^{-1}$ centered on 4740 Å. For the second run in October 2007, the SITe-5 CCD detector was used to image spectra at a resolution of 60 Å mm$^{-1}$ centered on 4200 Å. The spectra were reduced and analyzed using the NOAO's routines in IRAF along with software packages by Christian Buil (Iris), Valérie Desnoux (Vspec), and Robert H. Nelson (RaVereC). Lastly, a search for periodicity in the photometry of periodic variables in the sample was carried out in the Peranso software environment (Vanmunster 2007) using the algorithms ANOVA (Schwarzenberg-Czerny 1996), FALC (Harris *et al.* 1989), and CLEANest (Foster 1995).

## 3. Comments on Individual Stars

3.1 BD+66° 1673 (EA, O5 V((f))n)

BD+66° 1673 lies on the northwestern edge of Berkeley 59, an open cluster embedded in an H II region (Figure 1) at the core of the Cep OB4 association. The star was classified previously from objective-prism spectra as O9–B0 (MacConnell 1968; Walker 1965), but is reclassified as O5 V((f))n from a spectrum at 60 Å mm$^{-1}$ obtained at the DAO (Figure 2). BD+66° 1673 now has the distinction of being the hottest star in Cep OB4. The star's implied high temperature drives mass loss via strong stellar winds that may be largely responsible for shaping and illuminating the surrounding H II region, along with playing a role in the formation of clearly-visible, cold, molecular pillars (Yang and Fukui 1992; Gahm *et al.* 2006).

BD+66° 1673 was monitored as part of a search for variable stars in Berkeley 59, and was the first to exhibit convincing brightness variations. A period analysis of the photometry revealed a strong signal for P = 5.33146 ±0.02000 days. The phased light curve (Figure 3) is that of an Algol-type eclipsing binary system, with primary and secondary eclipse depths of ~0.13 and ~0.06 mag., although with eclipses skewed in phase indicating an eccentric orbit of $e \cong 0.3$ and longitude of periastron $\omega \cong 180°$. Our adopted ephemeris for light minimum in the system from regression fits for the light curve is:

$$\text{HJD}_{min} = 2454400.5322 + 5.33146\ E.$$

For the inferred stellar mass and temperature of the primary, a preliminary model generated for the system using *Binary Maker 3* (Bradstreet and Steelman 2004) constrains the companion to be an early-to-mid B-type star (B2–B3, say) in a system with an orbital inclination of ~75°. A more detailed photometric study is needed to generate a full light curve to confirm the orbital eccentricity as well as to permit a more detailed analysis.

The distance to BD+66° 1673 can be established from its likely membership in Berkeley 59 and Cep OB4, despite a spatial location outside the cluster nucleus. We obtained all-sky *BV* photometry of the cluster field on five nights, derived extinction coefficients using the techniques outlined by Henden and Kaitchuck (1998), and standardized the photometry to the Johnson system using stars in the nearby cluster NGC 225 (Hoag *et al*. 1961). Our data for cluster stars for which we have spectroscopic results are given in Table 2.

*BV* photometry does not permit one to assess the properties of the dust extinction in the field, and for that we must rely on the *UBV* photometry obtained by MacConnell (1968) for bright association members. A complete reanalysis is indicated, given that MacConnell (1968) derived different values for the ratio of total to selective extinction ($R_V$) for different stars and regions of Cep OB4. Unusual reddening properties for the dust in the field were first suggested by Blanco and Williams (1959) in their study of Cep OB4. Such properties, if confirmed, would limit the accuracy of any derived distance to the cluster and association stars. We therefore decided to reanalyze the $R_V$ estimate from the available photometric and spectroscopic data.

The new spectral classifications for Cep OB4 stars (Figure 2, Table 2) can be used with the *UBV* photometry of MacConnell (1968) to establish the reddening law for the dust in the field. The result for the four brightest stars is a reddening slope of $X = 0.80 \pm 0.01$, identical to the reddening parameters inferred for dust obscuring nearby fields in Cygnus (Turner 1989). The observed *UBV* colors were dereddened with that reddening law and either zero-age main-sequence (ZAMS) luminosities from Turner (1976a, 1979) or main-sequence luminosities for obvious non-ZAMS stars, with results presented in the variable-extinction diagram of Figure 4. Some stars have colors systematically too blue for their observed spectral types, a feature encountered for many early-type emission-line stars (Schild and Romanishin 1976). The reddening law in Cep OB4 is reasonably well defined by the data, and the slope of the relation depicted in Figure 4 is $R_V = 2.81 \pm 0.09$ from least squares and non-parametric straight line fits. The slope is consistent with the properties of nearby dust clouds (Turner 1996b).

The distance to Berkeley 59 and Cep OB4 follows from a ZAMS fit, which gives $V_0–M_V = 9.73 \pm 0.11$, corresponding formally to $d = 883 \pm 43$ pc. A main-sequence fit to our new photometry for stars in the core of Berkeley 59 (Table 2) is shown in Figure 5 along with data for association members. The results reveal another feature, namely an excess reddening for the B8 III star 2MASS 00021063+6724087: $E_{B–V} = 1.5$ relative to $E_{B–V} = 1.38 \pm 0.02$ for other, spatially-adjacent, cluster members. 2MASS 00021063+6724087 may be an example of a rotating star embedded in a dust torus (see Turner 1996a), and its location in Figure 5 is consistent with a star near the turn-on point for pre-main-sequence objects.

The cluster radial velocity can be deduced by merging observations from Crampton and Fisher (1974) and Liu *et al*. (1989, 1991), which yields $V_R = –7 \pm 15$ km s$^{-1}$ for BD+66° 1674, and $–8 \pm 31$ km s$^{-1}$ for BD+66° 1675. Crampton and Fisher (1974) noted that scatter in the radial velocities for both objects suggested they are spectroscopic binaries. A period search reveals putative periods of $P = 1.20$ days and $P = 5.30$ days for BD+66° 1674. The results are sufficiently interesting to justify additional radial velocity measures, which are essential to establish a unique solution.

The presence of a O5 V((f))n star in Berkeley 59 and the predominance of late B-type cluster members lying above the ZAMS indicates an extremely young cluster. An age of ~$2 \times 10^6$ years for Berkeley 59 is estimated from BD+66° 1673 and the luminosities of stars lying near the cluster turn-on point in Figure 5 (see Guetter and Turner 1997), marking Berkeley 59 as one of the youngest and nearest clusters known. Certainly O5 V((f))n stars are rarely encountered in our Galaxy within a kiloparsec of the Sun. Much like Berkeley 87 discussed later, Berkeley 59 contains a sufficient number of exotic stars to make it an object of intense interest for future studies.

3.2  2MASS 00104558+6127556 (EA, A9 V)

The star 2MASS 00104558+6127556 was found to vary in light during monitoring of the Cepheid BD Cas. A dominant period of $P = 2.7172 \pm 0.0060$ days produces a light curve (Figure 3) indicating an Algol-type eclipsing system with eclipse depths of ~0.43 mag. for primary minimum and ~0.31 mag. for secondary minimum. Our adopted ephemeris for light minimum in the system from regression fits for the light curve is:

$$HJD_{min} = 2454404.8586 + 2.7172 \, E.$$

A spectrum (Figure 6) indicates a spectral type near A9 V, but with some features that may indicate contamination by a companion. For the inferred stellar mass and temperature of the primary, a preliminary model generated for the system using Binary Maker 3 (Bradstreet and Steelman 2004) constrains the companion to be a mid F-type star (F4-F5, say) in a system with an orbital inclination of ~88°. Additional observations are needed to refine the period and to establish a complete light curve for a formal analysis.

3.3  2MASS 19064659+4401458 (XI?, G2 V)

2MASS 19064659+4401458 lies in the field of the nova-like cataclysmic variable MV Lyr, and was found to display a low level of variability while monitoring the suspected Cepheid NSV 11753 (see section 3.8).  The star's variability was presumably discovered by Weber (1959, number 90 in his list), although it is misidentified in the original finder chart as a different star of constant brightness located a mere 1.4 arcminutes away (NSV 11753, J2000 19:06:54.19, +44:02:55.5). There is X-ray emission at a flux level of $2.21 \times 10^{-2}$ counts s$^{-1}$ from within an arcminute of the object (ROSAT ASSC source J190645.9+440139, Voges *et al.* 2000) that is distinct from an X-ray flux of $1.26 \times 10^{-2}$ counts s$^{-1}$ associated with MV Lyr itself (ROSAT ASSC source J190716.8+440109, Voges *et al.* 1999). The object's spectral type is G2 V (Figure 6), and its spectroscopic parallax (below) implies an X-ray luminosity of $L_X \cong 10^{30}$ ergs s$^{-1}$, as estimated using the energy conversion factor of Huensch *et al.* (1996). The result is too low for a canonical low-mass X-ray binary ($L_X \geq 10^{36}$ ergs s$^{-1}$), but is similar to that of W UMa systems ($L_X \sim 10^{29-30}$ ergs s$^{-1}$, Chen *et al.* 2006), chromospherically active stars (*i.e.*, RS CVn variables), and compact binaries.

A preliminary period analysis resulted in a dominant signal corresponding to $P \cong 7.0$ days, which is consistent with the initial estimate by Weber (1959) of 8 days, and agrees with more recent estimates by David Boyd and Christopher Lloyd (private communication). The phased light curve in Figure 3 displays a brief eclipse superposed on more rapid, irregular, quasi-sinusoidal variations. The possibility that 2MASS 19064659+4401458 is a close contact system is inconsistent with the inferred period, and chromospheric activity in the primary is precluded by the absence of Ca II H and K emission in its spectrum.

An alternative scenario would associate irregularities in the light curve with frictional heating in an accretion disk, implying that 2MASS 19064659+4401458 is a semi-detached binary system consisting of a G dwarf overfilling its Roche surface and orbiting a compact companion, presumably a white dwarf. During eclipses the irregular variations disappear, implying that they originate from a hot spot in the accretion disk that is eclipsed by the G2 dwarf. It is hoped that the preliminary results presented here will motivate additional observers to join an ongoing campaign led by variable star observer David Boyd, with a group including Richard Huziak, Roger Pickard, Tomas Gomez, Gary Poyner, Tom Krajci, and Bart Staels, to constrain the period and further investigate the star. An understanding of the system via optical photometry, however, may be limited by the nature of the source driving the irregular variations. Time-series spectroscopy is also needed to assess the full nature of the system.

The object's distance is estimated from the canonical distance modulus relation reformulated for the infrared, namely with $A_J = 0.276 \times A_V$ and $E_{J-H} = 0.33 \times E_{B-V}$ (Bica and Bonatto 2005, Dutra *et al*. 2002). The 2MASS magnitudes for the star are $J = 11.275 \pm 0.020$, $H = 10.776 \pm 0.018$, and $K = 10.650 \pm 0.016$ (Cutri *et al*. 2003), and an absolute magnitude and intrinsic color can be established from the star's main-sequence spectral type as $M_J = 3.24 \pm 0.53$ and $(J–H)_0 \cong 0.28 \pm 0.06$, parameters deduced from a sample of $n \cong 30$ G2 V stars in the Hipparcos database (Perryman *et al*. 1997) with cited parallax uncertainties ≤0.7 mas. The implied intrinsic color agrees with a value of $(J–H)_0 \cong 0.38$ from Koornneef (1983), when converted to the 2MASS system with the appropriate transformation (Carpenter 2001). The resulting distance for $R_V = 3.1$ is $d \approx 310$ pc, while, if a nearly negligible field reddening is adopted, as advocated by a 2MASS color-color diagram of the region, it is $d \approx 400$ pc.

3.4  BD+22° 3792 (SRB, M6 III)

The variability of BD+22° 3792, of spectral type M6 III (Shi and Hu 1999, see Figure 6), was discovered by the ASAS survey (Pojamski *et al*. 2005). The semi-periodic nature of its photometric variations (Figure 7) and its spectral type are consistent with a Type B semi-regular variable. A Fourier analysis of our observations and those of ASAS-3 implies a possible period around 79 days. The star's spectral energy distribution displays far-infrared emission, indicating the presence of a warm dusty envelope surrounding the star, likely stemming from mass loss.

BD+22° 3792 is 12′ from the open cluster NGC 6823, but is not a member. The cluster's associated H II region is excited by numerous, young, reddened, OB stars, which are ~4-7 × $10^6$ years old at a distance of $d = 2.1 \pm 0.1$ kpc (Guetter 1992). The star's distance can be estimated by spectroscopic parallax using photometry taken from Massey *et al*. (1995) and a set of intrinsic parameters, $M_V = –0.3 \pm 0.7$ and $(B–V)_0 = 1.36 \pm 0.09$, derived from a sample of $n = 7$ nearby M6 giants in the Hipparcos database (Perryman *et al*. 1997) with cited parallax uncertainties ≤0.9 mas. The intrinsic parameters for M6 giants in the literature are rather scattered (see Mikami 1986, and references therein), mainly because most are variable and exhibit intrinsic color excesses. The resulting distance to BD+22° 3792 is ~700 pc for a reddening of $E_{B-V} \cong 0.38$ from the Hipparcos data, while adoption of an intrinsic color of $(B–V)_0 = 1.58$ from Lee (1970) gives $d \approx 950$ pc and $E_{B-V} \cong 0.18$.

### 3.5 2MASS 19475544+2722562 (SRB, M4 III)

The star 2MASS 19475544+2722562 lies in the field of the Cepheid S Vul. Its light-curve (Figure 7) exhibits a nearly regular oscillatory trend superimposed upon a gradual increase in brightness. The star's late spectral type of M4 III (Figure 6) suggests a likely designation as a Type B semiregular. A Fourier analysis suggests a rather short period of ~27 days for the oscillatory trend.

### 3.6 ALS 10588 = LS II+27 19 (SPB? ELL?, B3 IVn)

Alma Luminous Star 10588 (Reed 1998), or LS II+27 19 (Stock *et al*. 1960), is a close spatial neighbor of the Cepheid SV Vul, and a likely member of Vul OB1 (Turner 1980, 1984). The star's evolutionary status from its spectral type (B3 IVn, Figure 6) matches that of stars associated with SV Vul, namely main-sequence objects terminating at B3. Spectroscopic parallax places the star at a distance of 2040 ±470 pc, consistent with the distance to Vul OB1 ($d$ = 2.1–2.5 kpc, Turner 1986; Guetter 1992). ALS 10588 exhibits an IR excess with emission at 60 $\mu$m and 100 $\mu$m in the IRAS survey (IRAS19498+2717), which might account for its larger space reddening of $E_{B-V}$ = 0.79 ±0.02 relative to a value of $E_{B-V}$ = 0.50 for SV Vul, provided that the former possesses an equatorial dust torus (see Turner 1996a).

The variability of ALS 10588 was revealed during monitoring of SV Vul, although it also appears to have been detected in a *VI* survey for new Cepheids by Metzger and Schechter (1998). A period analysis of the photometry revealed a dominant signal at $P$ = 1.8521 ±0.0005 days, although a solution for twice that value ($P$ = 3.704 days) matches our observations and those of ASAS (Pojamski *et al*. 2005) (Figure 8). The spectral type and period are consistent with the class of slowly pulsating B stars (SPBs), characterized by stars of spectral types B3–B8 oscillating with periods on the order of days (Waelkens and Rufener 1985). The observed *V* amplitude ($\approx$ 0.25 mag.), however, is unusually large for a SPB variable (De Cat *et al*. 2000). Similarly, if twice the period is adopted, the inferred ellipsoidal system has a light amplitude more than twice that observed in other B stars of the same class (Beech 1989). There is also an absence of spectroscopic contamination from the expected companion (Figure 6). A toroidal dust clump orbiting synchronously with the star would account for the IR excess as well as the star's excess reddening (Turner 1996a), and, if tied to the star's rotation, would imply a stellar rotational velocity of ~250 km s$^{-1}$, consistent with the slightly diffuse nature of the star's spectrum. Yet there is no significant deviation from a repeatable light curve morphology over several seasons of observation. The star's status may ultimately need to be resolved by time-series spectroscopy to examine whether the resulting radial velocities are consistent with the trend noted for SPBs (De Cat and Aerts 2002), or the canonical features of binarity or extrinsic variability.

### 3.7 HDE 229059 (*α* Cyg variable, B2 Iabe)

HDE 229059 is a B2 Iabe supergiant that displays emission in the lower hydrogen Balmer lines (Figure 6) and has an infrared (IR) excess (Clarke et al. 2005). Such characteristics indicate active mass loss and the presence of circumstellar dust. The *General Catalogue of Variable Stars* (Samus *et al*. 2004) designates stars with comparable spectral types and *V* amplitudes ($\approx$ 0.1 mag, Figure 7) to those of HDE 229059 as *α* Cyg variables, with irregular light variations tied to overlapping modes of non-radial pulsation. Burki *et al.* (1978) suggest that all luminosity class Ia B–G supergiants probably vary in brightness (see also Bresolin *et al.* 2004, and references therein).

HDE 229059 lies in Berkeley 87, an open cluster that has received considerable attention because it is a strong source of $\gamma$ and cosmic rays (Giovannelli 2002; Aharonian *et al.* 2006), which has motivated an area of research on how stellar winds interact with the interstellar medium, enabling young open clusters to become pseudo particle accelerators. Berkeley 87 also hosts an abundance of astronomical phenomonae (compact H II regions, molecular clouds, masers, and radio sources) and exotic stellar constituents that includes V439 Cyg, Stephenson 3, and BC Cyg. V439 Cyg is an emission-line star that has exhibited dramatic spectroscopic changes over a short time-scale (Polcaro *et al.* 1990; Polcaro and Norci 1997; Norci *et al.* 1998; Polcaro and Norci 1998), Stephenson 3 is a rare type of Wolf-Rayet star (WO3) (Norci *et al.* 1998; Polcaro *et al.* 1997), and BC Cyg is an M3 Ia supergiant and type C semiregular variable (Turner *et al.* 2006) that will eventually terminate in a Type II supernova explosion. The cluster therefore offers intriguing insights into the effects of initial mass and mass loss on the end states of evolution for O-type stars, and may allow us to place new constraints on the initial progenitor masses for WO stars and red supergiants. The situation of HDE 229059 in such an evolutionary scheme is not entirely clear, which is why further study is essential. As a start, we investigate the possibility of cluster membership for the stars using spectroscopic parallax, 2MASS photometry, and proper motion data.

The distance to HDE 229059 can be established by spectroscopic parallax using the photometry of Turner and Forbes (1982) and intrinsic parameters determined from a sample of blue supergiants: $M_V = -6.4 \pm 0.8$ and $(B-V)_0 = -0.19 \pm 0.03$ (Kudritski *et al.* 1999; Blaha and Humphreys 1989; Garmany and Stencel 1992), values that compare favorably with unpublished results (Turner) of $M_V = -6.3$ and $(B-V)_0 = -0.18$ for B2 Iab stars. The distance for $R_V = 3.0$ is $d = 970$ pc. For BC Cyg, with mean $\langle V \rangle$ and $\langle B-V \rangle$ from Turner *et al.* (2006) and intrinsic parameters derived for the comparable M-type supergiant Alpha Orionis, the resulting distance is $d \approx 1200$ pc.

2MASS photometry (Cutri *et al.* 2003) for the cluster field yields color-color and color-magnitude diagrams for Berkeley 87 presented in Figure 9. The reddening solution, $E_{J-H} = 0.42 \pm 0.04$ ($E_{B-V} \cong 1.36$), is well-defined because of the presence of numerous young B-type stars in the cluster. Isochrones for the 2MASS system (Padova Database of Stellar Evolutionary Tracks and Isochrones, Bonatto *et al.* 2004) fit the data at a distance of $d = 1280 \pm 150$ pc. The reddening matches previous results, but the distance is larger than that found by Turner and Forbes (1982), although consistent with a later estimate of $1230 \pm 40$ pc (Turner *et al.* 2006). Constraining the cluster's age from 2MASS data is complicated by the fact that BC Cyg lies near the saturation limit of the survey. The isochrone fit in Figure 9 is provided mainly to highlight the envisioned evolution, although high mass loss rates are indicated and the plotted isochrone is more closely linked with conservative mass evolution.

Proper motion data (Zacharias *et al.* 2004) exist for several bright stars whose membership in Berkeley 87 is supported by their locations in Figure 9, and can be compared with the similar values found for HDE 229059, BC Cyg, and Stephenson 3 (Table 3). The proper motions for such distant stars are small and may be dominated by measuring uncertainties. Consequently, we can only argue that a physical association between the above stars and the cluster cannot be excluded on the available evidence. Membership of the exotic variable stars of Berkeley 87 would be strengthened by radial velocity measures.

3.8 NSV Variables

A number of stars from the *Catalogue of Newly Suspected Variable Stars* (NSV, Samus *et al*. 2004) were surveyed in a search for potential small-amplitude Cepheids. Reference stars of well-established magnitude in each field were not available, so the observations were made differentially relative to other stars in the field, with unknown zero-point. The co-ordinates provided by the original sources are estimated to be uncertain by several arcminutes or more, which led us to make photometric sweeps of the immediate field to find objects that might correspond to the suspected variables. There are stars that are reasonably coincident with the co-ordinates for the NSV variables listed in Table 4, but none appear to be light variable. Listed in Table 4 are the co-ordinates from the NSV for the suspected variables, magnitudes from Samus *et al*. (2004), the standard deviation of the magnitude estimates for the stars selected in the present survey, and the number of observations made. None of the stars identified here in the fields of the suspected Cepheid variables displayed the canonical light variations expected, although other types of variability cannot be dismissed because of our limited observational sampling.

**4. Discussion**

It is of interest to note how a program of regular observation of Cepheid variables has generated serendipitous discoveries of new variable stars because of the need to establish photometric standard stars and check stars in the fields of the CCD images. In many cases the variable stars prove to be interesting, possibly unique, objects in their own right. But additional photometric and spectroscopic observations may be essential for clarifying their overall properties.

We are indebted to the following groups for facilitating the research described here: the staff at la Centre de Données astronomiques de Strasbourg, 2MASS, and NASA's Astrophysics Data System (ADS). We are particularly grateful to Conny Aerts for relevant discussions on SPBs, David Boyd and Christopher Lloyd for sharing their insights on 2MASS 19064659+4401458, Robert H. Nelson for sharing his expertise in various areas, and Dmitry Monin, Les Saddelmeyer, and the rest of the staff at the Dominion Astrophysical Observatory.

**References**

Aharonian, F., et al. 2006, *Astron. Astrophys*., **454**, 775.
Beech, M. 1989, *Astrophys. Space Sci*., **152**, 329.
Blaha, C., and Humphreys, R. M. 1989, *Astron. J*., **98**, 1598.
Blanco, V. M., and Williams, A. D. 1959, *Astrophys. J*., **130**, 482.
Bica, E., and Bonatto, C. 2005, *Astron. Astrophys*., **443**, 465.
Bonatto, C., Bica, E., and Girardi, L. 2004, *Astron. Astrophys*., **415**, 571.
Bonatto, C., Bica, E., Ortolani, S., and Barbuy, B. 2006, *Astron. Astrophys*., **453**, 121.
Bradstreet, D. H., and Steelman, D. P. 2004, Binary Maker 3 Light Curve Synthesis Program (Contact Software: Norristown, Pennsylvania).

Bresolin, F., Pietrzyński, G., Gieren, W., Kudritzki, R.-P., Przybilla, N., and Fouqué, P. 2004, *Astrophys. J.*, **600**, 182.
Burki, G., Maeder, A., and Rufener, F. 1978, *Astron. Astrophys.*, **65**, 363.
Carpenter, J. M. 2001, *Astron. J.*, **121**, 2851.
Chen, W. P., Sanchawala, K., and Chiu, M. C. 2006, *Astron. J.*, **131**, 990.
Clarke, A. J., Oudmaijer, R. D., and Lumsden, S. L. 2005, *Monthly Notices Roy. Astron. Soc.*, **363**, 1111.
Crampton, D., and Fisher, W. A. 1974, *Publ. Dom. Astrophys. Obs. Victoria*, **14**, 283.
Cruz-González, C., Recillas-Cruz, E., Costero, R., Peimbert, M., and Torres-Peimbert, S. 1974, *Rev. Mex. Astron. Astrof.*, **1**, 211.
Cutri, R. M., Skrutskie, M. F., van Dyk, S., Beichman, C. A., Carpenter, J. M., Chester, T., Cambresy, L., Evans, T., Fowler, J., Gizis, J., Howard, E., Huchra, J., Jarrett, T., Kopan, E. L., Kirkpatrick, J. D., Light, R. M., Marsh, K. A., McCallon, H., Schneider, S., Stiening, R., Sykes, M., Weinberg, M., Wheaton, W. A., Wheelock, S., and Zacarias, N. 2003, *The IRSA 2MASS All-Sky Point Source Catalog of Point Sources*, NASA/IPAC Infrared Science Archive.
De Cat, P., Aerts, C., De Ridder, J., Kolenberg, K., Meeus, G., and Decin, L. 2000, *Astron. Astrophys.*, **355**, 1015.
De Cat, P., and Aerts, C. 2002, *Astron. Astrophys.*, **393**, 965.
Denny, R. 2007, DC-3 Dreams, SP, PinPoint Astrometric Engine software.
Dutra, C. M., Santiago, B. X., and Bica, E. 2002, *Astron. Astrophys.*, **381**, 219.
Egan, M. P., Price, S. D., and Kraemer, K. E. 2003, *Bull. Am. Astron. Soc.*, **35**, 1301.
Foster, G. 1995, *Astron. J.*, **109**, 1889.
Gahm, G. F., Carlqvist, P., Johansson, L. E. B., and Nikolić, S. 2006, *Astron. Astrophys.*, **454**, 201.
Garmany, C. D., and Stencel, R. E. 1992, *Astron. Astrophys. Suppl.*, **94**, 211.
George, D. 2007, MaxIm DL/CCD software, http://www.cyanogen.com.
Giovannelli, F. 2002, *Mem. Soc. Astron. Ital.*, **73**, 920.
Guetter, H. H. 1992, *Astron. J.*, **103**, 197.
Guetter, H. H., and Turner, D. G. 1997, *Astron. J.*, **113**, 2116.
Harris, A. W., Young, J. W., Bowell, E., Martin, L. J., Millis, R. L., Poutanen, M., Scaltriti, F., Zappala, V., Schober, H. J., Debehogne, H., and Zeigler, K. W. 1989, *Icarus*, **77**, 171.
Helou, G., and Walker, D. W. 1988, *Infrared Astronomical Satellite (IRAS) Catalogs and Atlases*. Volume **7**, p.1-265, 7.
Henden, A. A., and Honeycutt, R. K. 1995, *Publ. Astron. Soc. Pacific*, **107**, 324.
Henden, A. A., and Kaitchuck, R. H. 1998, *Astronomical Photometry: A Text and Handbook for the Advanced Amateur and Professional Astronomer* (Willmann-Bell, Inc.: Richmond, Virginia).
Hoag, A. A., Johnson, H. L., Iriarte, B., Mitchell, R. I., Hallam, K. L., and Sharpless, S. 1961, *Publ. U. S. Naval Obs., Second Series*, **17**, 343.
Høg, E., Fabricius, C., Makarov, V. V., Urban, S., Corbin, T., Wycoff, G., Bastian, U., Schwekendiek, P., and Wicenic, A. 2000, *Astron. Astrophys.*, **355**, L27.
Huensch, M., Schmitt, J. H. M. M., Schroeder, K.-P., and Reimers, D. 1996, *Astron. Astrophys.*, **310**, 801.
Kohoutek, L., and Wehmeyer, R. 1999, *Astron. Astrophys. Suppl.*, **134**, 255.
Koornneef, J. 1983, *Astron. Astrophys.*, **128**, 84.

Kudritzki, R. P., Puls, J., Lennon, D. J., Venn, K. A., Reetz, J., Najarro, F., McCarthy, J. K., and Herrero, A. 1999, *Astron. Astrophys.*, **350**, 970.
Kwok, S., Volk, K., and Bidelman, W. P. 1997, *Astrophys. J. Suppl.*, **112**, 557.
Lane, D. J. 2007, Abbey Ridge Observatory and Abbey Ridge Auto-Pilot software, http://www.davelane.ca/aro.
Lee, T. A. 1970, *Astrophys. J.*, **162**, 217.
Liu, T., Janes, K. A., and Bania, T. M. 1989, *Astron. J.*, **98**, 626.
Liu, T., Janes, K. A., and Bania, T. M. 1991, *Astron. J.*, **102**, 1103.
MacConnell, D. J. 1968, *Astrophys. J. Suppl.*, **16**, 275.
Massey, P., Johnson, K. E., and Degioia-Eastwood, K. 1995, *Astrophys. J.*, **454**, 151.
Metzger, M. R., and Schechter, P. L. 1998, *Astron. J.*, **116**, 469.
Mikami, T. 1986, *Astrophys. Space Sci.*, **119**, 65.
Neckel, T. and Klare, G. 1980, *Astron. Astrophys. Suppl.*, **42**, 251.
Norci, L., Meurs, E. J. A., Polcaro, V. F., Viotti, R., and Rossi, C. 1997, *Astrophys. Space Sci.*, **255**, 197.
Norci, L., Polcaro, V. F., Rossi, C., and Viotti, R. 1998, *Irish Astron. J.*, **25**, 43.
Perryman, M. A. C., Lindegren, L., Kovalevsky, J., Høg, E., Bastian, U., Bernacca, P. L., Crézé, M., Donati, F., Grenon, M., van Leeuwen, F., van der Marel, H., Mignard, F., Murray, C. A., Le Poole, R. S., Schrijver, H., Turon, C., Arenou, F., Froeschlé, M., and Petersen, C. S. 1997, *The Hipparcos and Tycho Catalogues*, ESA SP-1200 (ESA: Noordwijk).
Polcaro, V. F., Rossi, C., Norci, L., and Giovannelli, F. 1990, *Astrophys. Space Sci.*, **169**, 31.
Polcaro, V. F., and Norci, L. 1997, *Astrophys. Space Sci.*, **251**, 343.
Polcaro, V. F., and Norci, L. 1998, *Astron. Astrophys.*, **339**, 75.
Polcaro, V. F., Viotti, R., Rossi, C., and Norci, L. 1997, *Astron. Astrophys.*, **325**, 178.
Pojmanski, G., Pilecki, B., and Szczygiel, D. 2005, *Acta Astron.*, **55**, 275.
Reed, B. C. 1998, *Astrophys. J. Suppl.*, **115**, 271.
Ritter, H., and Kolb, U. 2003, *Astron. Astrophys.*, **404**, 301.
Samus, N. N., Durlevich, O. V., et al. 2004, *Combined General Catalogue of Variable Stars*, VizieR Online Data Catalog, II/250.
Schild, R., and Romanishin, W. 1976, *Astrophys. J.*, **204**, 493.
Schwarzenberg-Czerny, A. 1996, *Astrophys. J.*, **460**, L107.
Shi, H. M., and Hu, J. Y. 1999, *Astron. Astrophys. Suppl.*, **136**, 313.
Stock, J., Nassau, J. J., and Stephenson, C. B. 1960, *Luminous Stars in the Northern Milky Way, II* (Hamburger Sternwarte and Warner and Swasey Observatory: Hamburg-Bergedorf)
Turner, D. G. 1976a, *Astron. J.*, **81**, 97.
Turner, D. G. 1976b, *Astron. J.*, **81**, 1125.
Turner, D. G. 1979, *Publ. Astron. Soc. Pacific*, **91**, 642.
Turner, D. G. 1980, *Astrophys. J.*, **235**, 146.
Turner, D. G. 1984, *J. Roy. Astron. Soc. Can.*, **78**, 229.
Turner, D. G. 1986, *Astron. Astrophys.*, **167**, 157.
Turner, D. G. 1989, *Astron. J.*, **98**, 2300.
Turner, D. G. 1996a, in *The Origins, Evolutions, and Destinies of Binary Stars in Clusters, Astron. Soc. Pacific Conf. Series*, **90**, eds. E. F. Milone and J.-C. Mermilliod, p. 382.
Turner, D. G. 1996b, in *The Origins, Evolutions, and Destinies of Binary Stars in Clusters, Astron. Soc. Pacific Conf. Series*, **90**, eds. E. F. Milone and J.-C. Mermilliod, p. 443.
Turner, D. G. 1998, *J. Am. Assoc. Var. Star Obs.*, **26**, 101.

Turner, D. G., and Forbes, D. 1982, *Publ. Astron. Soc. Pacific*, **94**, 789.
Turner, D. G., and Burke, J. F. 2002, *Astron. J.*, **124**, 2931.
Turner, D. G., Horsford, A. J., and MacMillan, J. D. 1999, *J. Am. Assoc. Var. Star Obs.*, **27**, 5.
Turner, D. G., Rohanizadegan, M., Berdnikov, L. N., and Pastukhova, E. N. 2006, *Publ. Astron. Soc. Pacific*, **118**, 1533.
Vanmunster, T. 2007, Peranso Light Curve and Period Analysis Software, http://www.peranso.com.
Voges, W., et al. 1999, *Astron. Astrophys.*, **349**, 389.
Voges, W., et al. 2000, *IAU Circ.*, **7432**, 1.
Waelkens, C., and Rufener, F. 1985, *Astron. Astrophys.*, **152**, 6.
Walker, G. A. H. 1965, *Astrophys. J.*, **141**, 660.
Weber, R. 1959, *J. Obs.*, **42**, 106.
Yang, J., and Fukui, Y. 1992, *Astrophys. J.*, **386**, 618.
Zacharias, N., Monet, D. G., Levine, S. E., Urban, S. E., Gaume, R., and Wycoff, G. L. 2004, *Bull. Am. Astron. Soc.*, **36**, 1418.

Table 1. Monitored variable stars

| Star | RA(2000) | DEC(2000) | Type | P (days) | Sp. Type |
|---|---|---|---|---|---|
| BD+66° 1673 | 00:01:46.86 | +67:30:25.1 | EB | 5.33146 | O5 V((f))n |
| 2MASS 00104558+6127556 | 00:10:45.58 | +61:27:55.6 | EB | 2.7172 | A9 V |
| 2MASS 19064659+4401458 | 19:06:46.82 | +44:01:46.5 | XI? | 7.0: | G2 V |
| BD+22° 3792 | 19:43:53.00 | +23:11:36.0 | SRB | 79.4: | M6 III |
| 2MASS 19475544+2722562 | 19:47:55.52 | +27:22:57.8 | SRB | 27.3: | M4 III |
| ALS 10588 | 19:51:52.87 | +27:25:00.1 | SPB? | 1.8521 | B3 IVn |
| … | … | … | ELL? | 3.704 | … |
| HDE 229059 | 20:21:15.45 | +37:24:35.2 | α Cyg | … | B2 Iabe |

Table 2. Photometry and spectroscopy of Berkeley 59 members

| Star | MacC[1] | RA(2000) | DEC(2000) | V | B–V | Sp. Type |
|---|---|---|---|---|---|---|
| BD+66° 1673 | 3 | 00:01:46.91 | +67:30:24.3 | 10.07±0.04 | 1.31±0.03 | O5 V((f))n |
| BD+66° 1675 | 14 | 00:02:10.32 | +67:24:32.5 | 9.08±0.03 | 1.08±0.02 | O7 V |
| BD+66° 1674 | 13 | 00:02:10.68 | +67:25:44.5 | 9.60±0.04 | 1.07±0.02 | B0 IIIn |
| MacConnell 15 | 15 | 00:02:13.42 | +67:25:05.5 | 11.30±0.03 | 1.08±0.02 | B0.5 Vn |
| 2MASS 00020012+6725109 | A3 | 00:02:00.17 | +67:25:11.2 | 12.81±0.03 | 1.20±0.01 | B3 V |
| 2MASS 00021063+6724087 | … | 00:02:10.63 | +67:24:08.7 | 13.43±0.02 | 1.36±0.03 | B8 III |

Note: [1] Numbering from MacConnell (1968).

Table 3. Proper motion data for Berkeley 87 stars.

| Star[2] | Identity | $\mu_{RA}$ (mas/yr) | $\mu_{DEC}$ (mas/yr) |
|---|---|---|---|
| 3 | HDE 229059 | −4.5 ±0.6 | −5.3 ±0.7 |
| 4 | … | −5.2 ±0.7 | −5.5 ±1.0 |
| 13 | … | −5.6 ±0.7 | −7.5 ±0.7 |
| 15 | V439 Cyg | +1.1 ±5.4 | +11.2 ±5.4 |
| 25 | … | −3.9 ±0.7 | −5.7 ±1.1 |
| 26 | … | −5.7 ±1.3 | −4.1 ±2.4 |
| 29 | Stephenson 3 | −8.0 ±5.4 | −2.8 ±5.4 |
| 32 | … | −5.1 ±2.0 | −3.1 ±0.7 |
| 78 | BC Cyg | −4.5 ±1.1 | −7.7 ±1.1 |

Note: [2] Numbering from Turner and Forbes (1982).

Table 4. Cepheid Candidates from the NSV Catalogue.

| Star | RA(2000) | DEC(2000) | Mag.[3] | s.d. | n |
|---|---|---|---|---|---|
| NSV 00924 | 02:48:19.91 | +58:41:44.8 | 12.50 | ±0.003 | 3 |
| NSV 11753 | 19:06:54.19 | +44:02:55.5 | 13.50 | ±0.007 | 16 |
| NSV 11931 | 19:21:11.73 | +00:07:02.6 | 14.20 | ±0.014 | 7 |
| NSV 14094 | 22:16:16.71 | +49:13:13.8 | 12.10 | ±0.008 | 8 |
| NSV 14237 | 22:35:04.02 | +63:47:37.6 | 12.30 | ±0.006 | 8 |

Note: [3] From Samus et al. (2004).

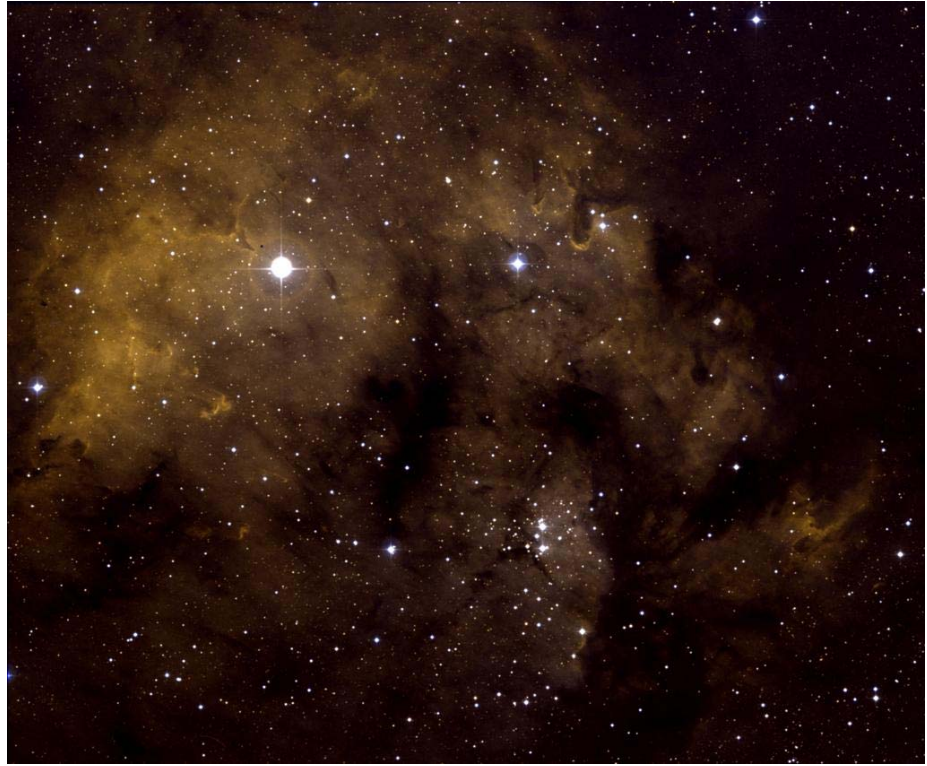

Figure 1. The field of view of Berkeley 59, a pseudo color image constructed from POSS II data.

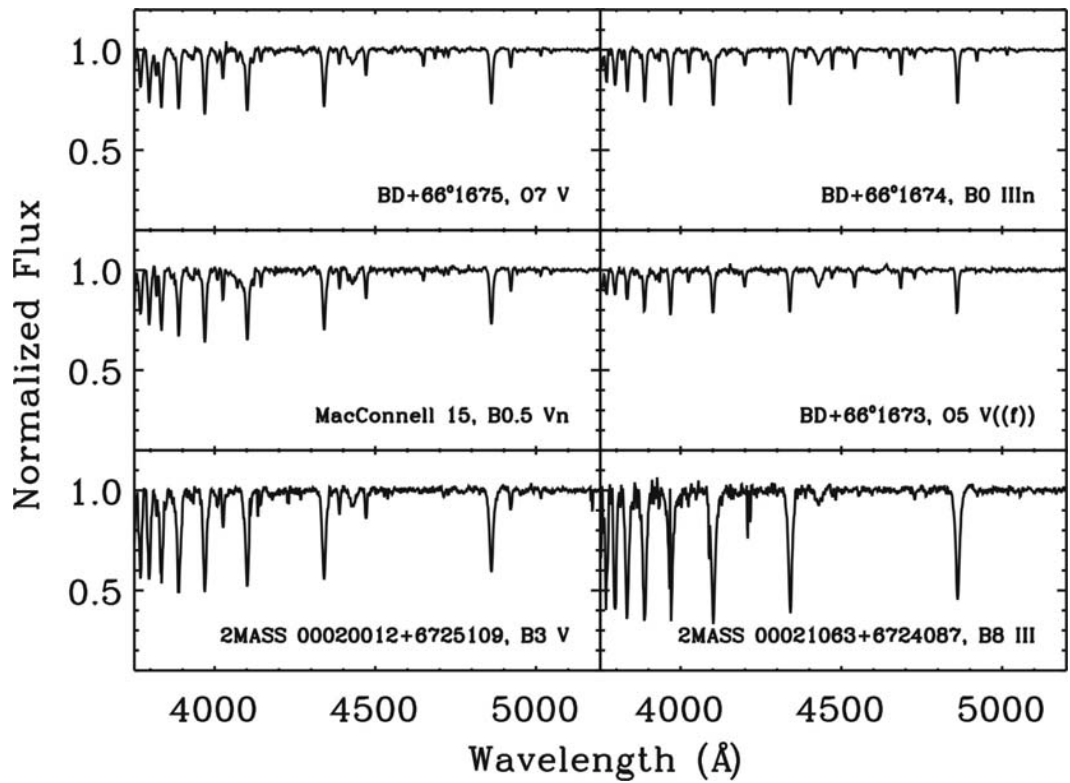

Figure 2. A mosaic of continuum-normalized CCD spectra for likely members of Berkeley 59 and Cep OB4.

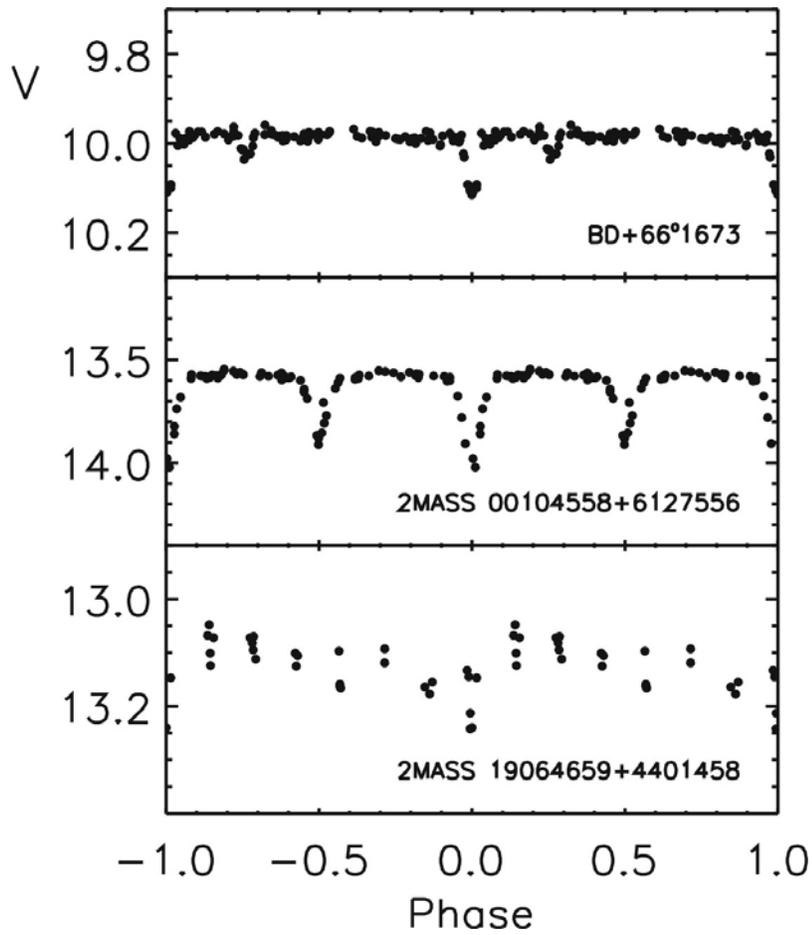

Figure 3. Light curves for the three eclipsing systems BD+66° 1673, 2MASS 00104558+6127556, and 2MASS 19064659+4401458.

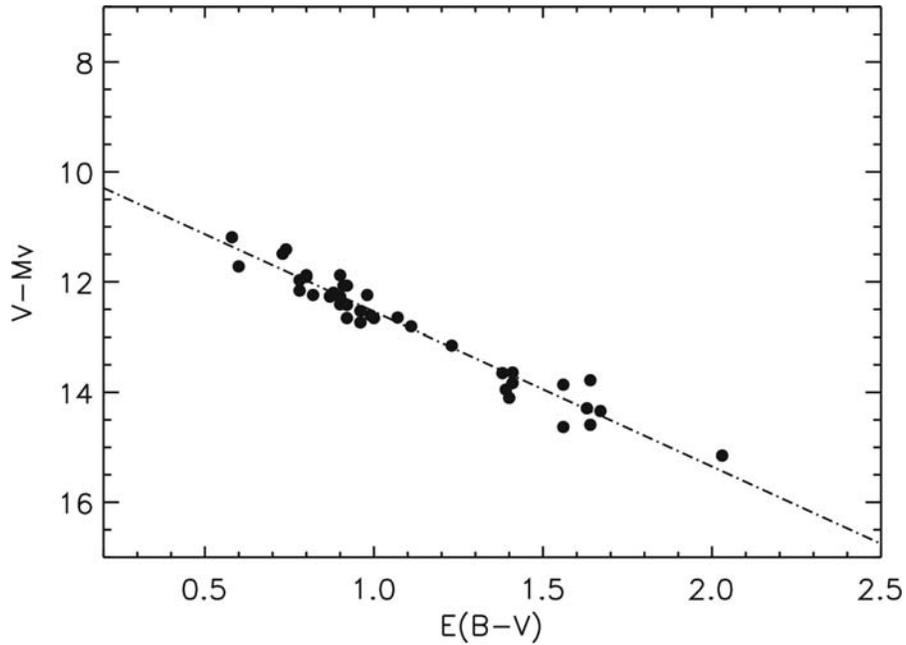

Figure 4. A variable-extinction diagram for likely main-sequence and zero-age main-sequence (ZAMS) members of Berkeley 59 and the Cep OB4 association. Least squares and non-parametric fits yield a ratio of the total to selective extinction of $R_V = 2.81 \pm 0.09$.

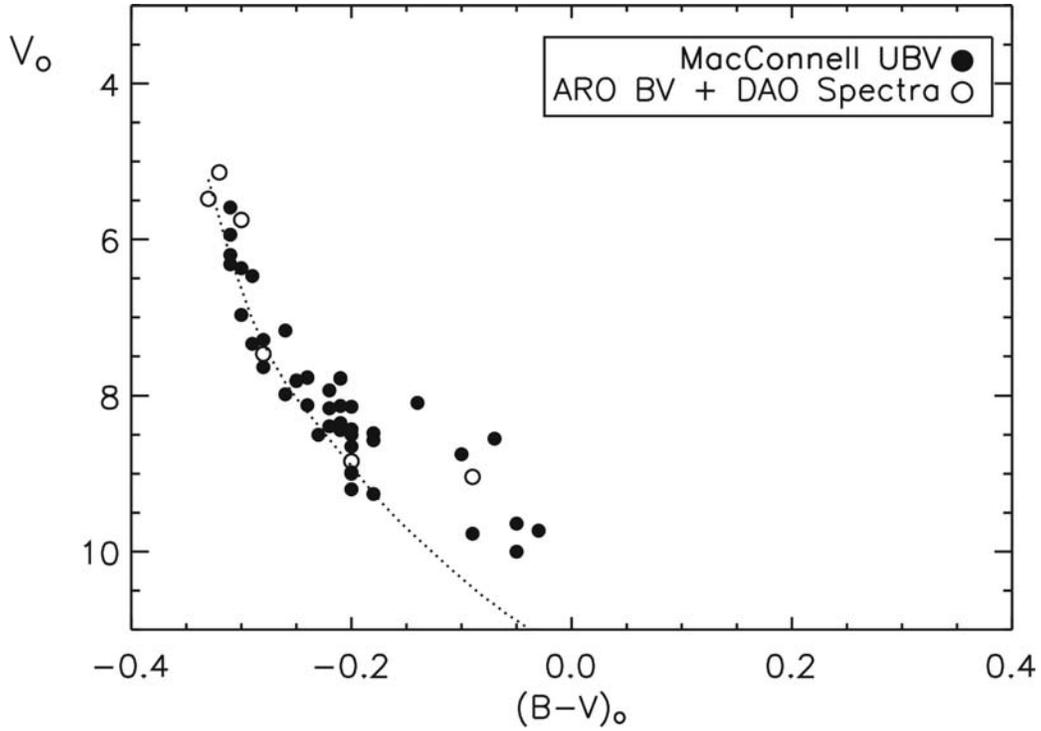

Figure 5. A reddening-free *BV* color-magnitude diagram for Berkeley 59 (open circles) and Cep OB4 (filled circles). A ZAMS fit to the observations yields a distance of $d = 883 \pm 43$ pc and a reddening of $E_{B-V} = 1.38 \pm 0.02$ in the core of the cluster.

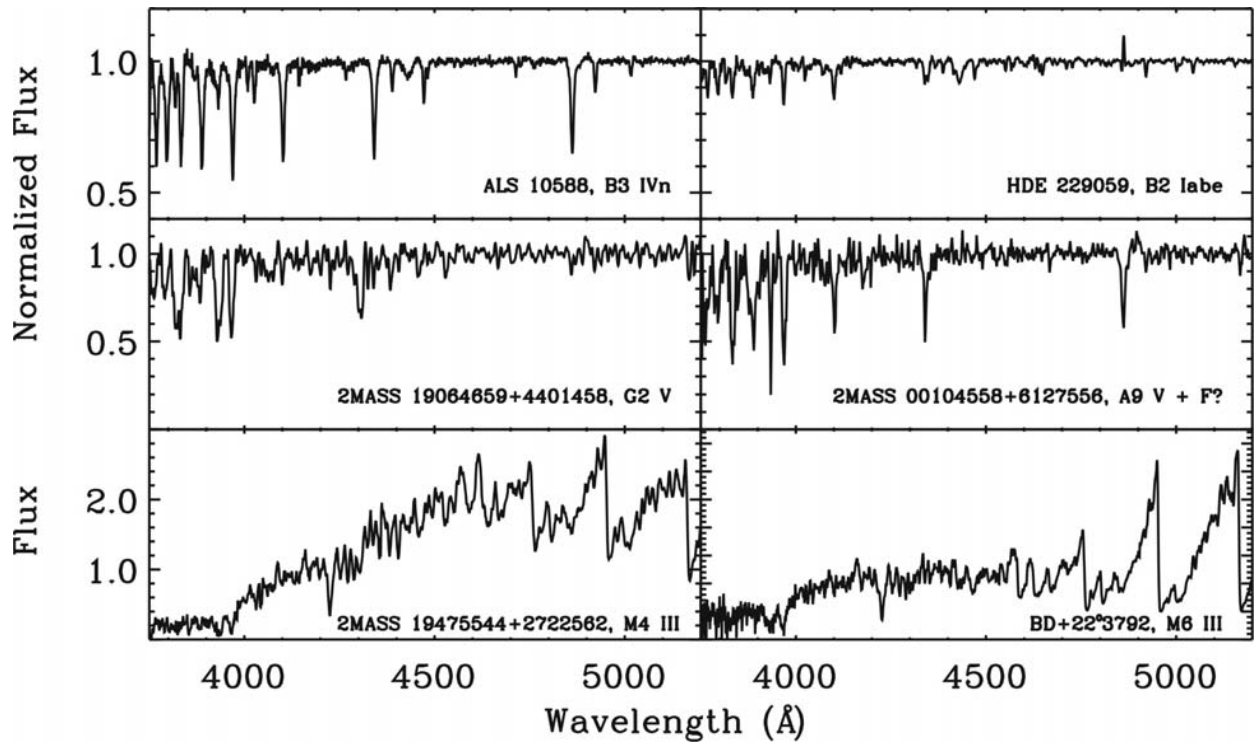

Figure 6. A mosaic of CCD spectra from the DAO 1.8-m Plaskett telescope for variables examined in this study.

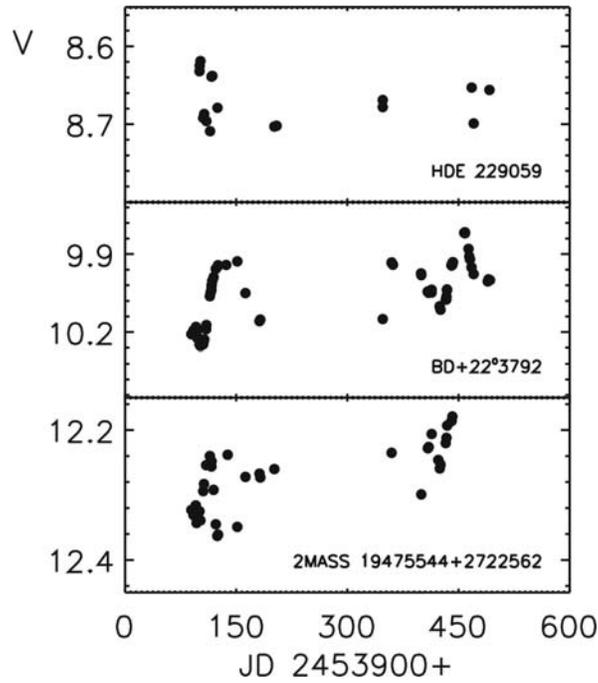

Figure 7. Light curves for variables examined in this study, constructed from differential photometry from the ARO. Zero-point offsets are expected (see text), although the standard deviation of observations for check stars in the same fields ranges from ±0.006 to ±0.008 mag.

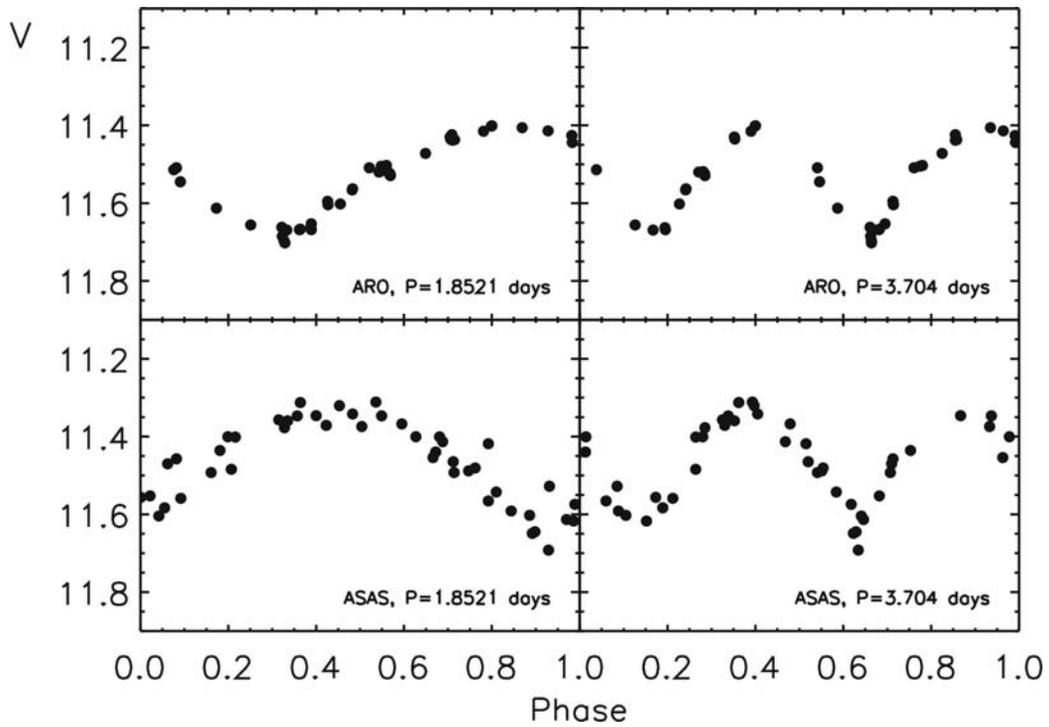

Figure 8. Light curves for ALS 10588 from ARO and ASAS data phased with possible periods of 1.8521 and 3.704 days.

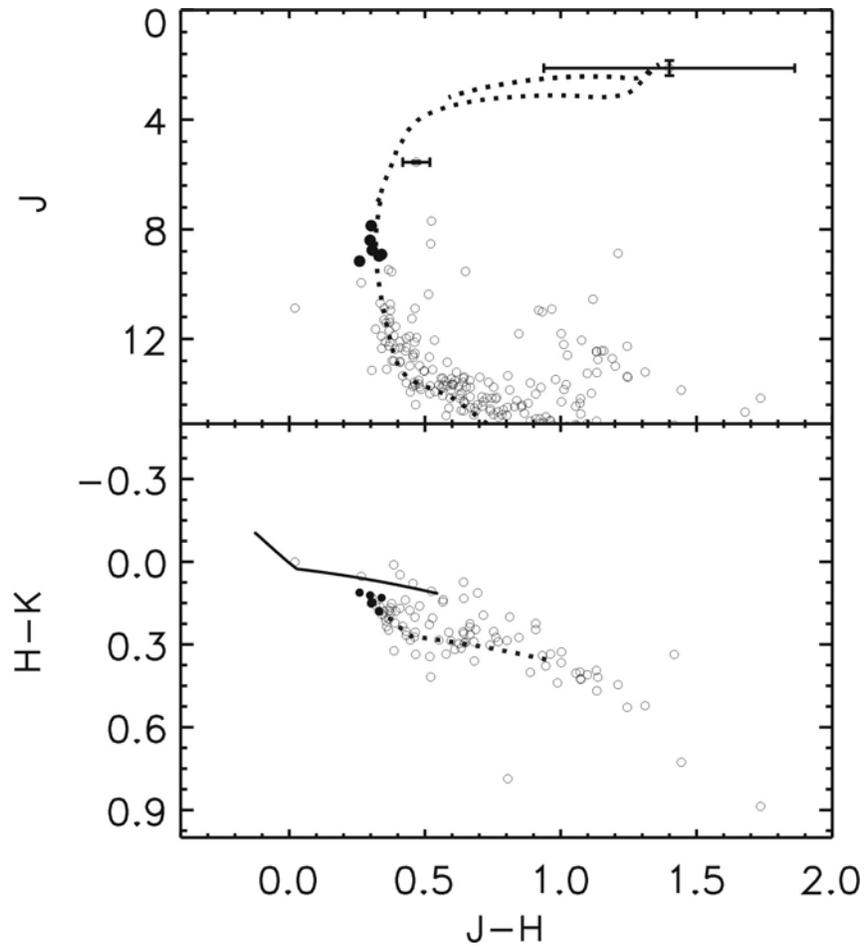

Figure 9. A color-color diagram (lower) and color-magnitude diagram (upper) for Berkeley 87 constructed from 2MASS data. The intrinsic color-color relation for the 2MASS system (Turner unpublished) is depicted as a solid line, as well as reddened by $E_{J-H} = 0.42 \pm 0.04$ ($E_{B-V} \cong 1.36$) as a dotted line (lower). Filled circles correspond to stars likely to be cluster members. An isochrone fit (upper) is provided to highlight the assumed evolution (see text). The variable stars HDE 229059 ($J = 5.551$) and BC Cyg ($J = 2.117$) are provided with photometric error bars (BC Cyg is near saturation).